\documentclass[runningheads]{llncs}

\usepackage{url}
\usepackage{enumitem}
\usepackage{multirow, graphicx}
\usepackage[dvipsnames]{xcolor}
\usepackage{booktabs,tabularx}
\usepackage{lmodern}
\usepackage[T1]{fontenc}
\usepackage[normalem]{ulem}
\newcommand\orangesout{\bgroup\markoverwith{\textcolor{orange}{\rule[0.5ex]{2pt}{0.4pt}}}\ULon}
\usepackage{framed}

\begin{document}
\sloppy
\title{A survey on test practitioners' awareness of test smells
%\title{Does your test smell? An Exploratory Survey on developers practices
% \thanks{Supported by organization x.}
}
\author{Nildo Silva Junior \and
Larissa Rocha \and Luana Almeida Martins \and
Ivan Machado}
\authorrunning{Silva Junior et al.}
\institute{Federal University of Bahia - UFBA, Salvador - BA, Brazil \\ \email{nildo.silva@ufba.br, larissars@dcc.ufba.br, martins.luana@ufba.br, ivan.machado@ufba.br} }

\maketitle              % typeset the header of the contribution

\begin{abstract}
%The abstract should briefly summarize the contents of the paper in 150--250 words.

Developing test code may be a time-consuming task that usually requires much effort and cost, especially when it is done manually. Besides, during this process, developers and testers are likely to adopt bad design choices, which may lead to the introduction of the so-called test smells in test code. Test smells are bad solutions to either implement or design test code. As the test code with test smells increases in size, these tests might become more complex, and as a consequence, much harder to understand and evolve correctly. Therefore, test smells may have a negative impact on the quality and maintenance of test code and may also harm the whole software testing activities. In this context, this study aims to understand whether test professionals non-intentionally insert test smells. We carried out an expert survey to analyze the usage frequency of a set of test smells. Sixty professionals from different companies participated in the survey. We selected 14 widely studied smells from the literature, which are also implemented in existing test smell detection tools. The yielded results indicate that experienced professionals introduce test smells during their daily programming tasks, even when they are using standardized practices from their companies, and not only for their personal assumptions. Another relevant evidence was that developers' professional experience can not be considered as a root-cause for the insertion of test smells in test code.

\keywords{Test smells \and Empirical study \and Expert Survey.}
\end{abstract}
\section{Introduction}

%contexto

Software projects, both commercial and open source, commonly encompass a set of automated test suites, as a crucial support to verify software quality \cite{Garousi2016STCE}. However, creating test code may require high effort and cost \cite{Garousi2016STCE,yusifouglu2015mapping}. For example, the SCADA project used 870 hours to create a set of test scripts \cite{wiederseiner2010open}. Automated test generation tools, such as Randoop\footnote{\url{https://randoop.github.io/randoop/}}, JWalk\footnote{\url{http://staffwww.   dcs.shef.ac.uk/people/A.Simons/jwalk/}}, and Evosuite\footnote{\url{http://www.evosuite.org/}}, emerge as alternatives to facilitate and streamline this process. 
If designed with high quality, automated testing offers benefits over manual testing, such as repeatability, predictability, efficient test runs, and thereby require less effort and costs \cite{Garousi2018mapping,yusifouglu2015mapping}. In this way, tests should be concise, repeatable, robust, sufficient, necessary, clear, efficient, specific, independent, maintainable, and traceable \cite{Meszaros2003}. %However, while those tools automate tests, they might not guarantee the quality of tests} \cite{Palomba2016testsmells,virginio2019coverage. 

However, the development of a well-designed test code is not a straightforward task. Developers are usually under time pressure and must deal with constrained budgets, which can stimulate the use of anti-patterns in test code, leading to the occurrence of the so-called test smells. Test smells are identifiers of poor implementation solutions and problems in test code design \cite{Greiler2013testsmells}. As a consequence of introducing test smells, %it is likely that the produced test code would 
the produced test code would likely have reduced quality and, consequently, may not reach its expected capabilities at finding bugs while being understandable, maintainable, etc. \cite{Garousi2018mapping,yusifouglu2015mapping}. The literature reports %a number of 
196 test smell types, classified in the following groups: behavior, logic, design related, issue in test steps, mock and stub-related, association in production code, code-related, and dependencies \cite{Garousi2018mapping}. %Test smells are present in several projects, reaching 82\% of JUnit tests \cite{Bavota2012smells}, such as \texttt{Assertion Roulette, Mystery Guest and Resource Optimism}. These smells may make it difficult to identify the statement that failed, make it difficult to understand the test due to unknown values, or may not determine the result, depending on the state of resources, for example. In this way, software developers may leave test activities because they became complex and hard to maintain \cite{Meszaros2003}. 
    
%\orangesout{Meszaros et al. \cite{Meszaros2003} define a set of  characteristics for automated test case. According to them, software testing should be concise, self checking, repeatable, robust, sufficient, necessary, clear, efficient, specific, independent, maintainable, and traceable. The detection of test smell is an alternative to evaluate the test code quality.}

Some studies have been conducted to identify and analyze the effect of the presence of test smells in software projects in several respects \cite{Garousi2016STCE,Greiler2013testsmells,palomba2019,Rompaey2006ArgoUML}.
%\and check software code coverage \cite{smeets2009automated} and \cite{virginio2019coverage}.  
In those studies, the test smells were presented as non-functional quality attributes within the Software Test Code Engineering process, for example. Some test smell types are presented, and there is a discussion on what they may cause in the test code maintenance \cite{Garousi2016STCE}. Metrics were defined to identify test smells on automated tools and these tools were validated by developers \cite{Greiler2013testsmells} or through a case study \cite{Rompaey2006ArgoUML}. Some test smells were also analyzed as a way to reduce flaky tests, i.e. tests with non-deterministic behavior \cite{palomba2019}. 
However, discussions about daily practices and programming styles that may contribute to inserting test smells are still lacking.
% However, it is still missing discussing around daily practices and programming vices that may contribute to insert test smells.
Understanding the relationship between development practices and test smell insertion may support improving the test creation process.
% During test code quality verification process, 
%Many studies have been conducted to identify test smells \cite{Greiler2013testsmells}, \cite{Garousi2016STCE}, \cite{Rompaey2006ArgoUML} and check code coverage \cite{smeets2009automated} and \cite{virginio2019coverage}. 
%Those studies present smells as non-functional quality attributes within the Software Test Code Engineering process.; or also as metrics for automatic smells detection to compare code coverage between tests created from tools and manual tests. However, it is still missing the discussion around daily practices and programming vices that may contribute to generate test smells. 
    
% With this information it is possible to perform an improvement in the quality of the creation of the test codes, either with a change in the code of test generation tools or with training of professionals. This would contribute to the reduction of smells generation and consequently the need to refactor the test code.

%objetivo
    
%This study aims to analyze practices that may introduce test smells and the frequency of these practices are used in the creation process of test code and the presence of these practices on tests during test execution. 
The %\textcolor{red}{\sout{This}} 
study aims to understand whether professionals non-intentionally insert test smells. Through an expert survey \cite{Linaker:TR:2015}, we could analyze the practices adopted by developers that might introduce test smells and how frequently these practices are used during test creation and observed during test execution. %Besides, we analyze the relation between professional experience and the use of test smells.
%This empirical study was performed based on a qualitative survey with 
The survey counted on sixty participants %\textcolor{blue}{from officially 8 companies and direct messages,} 
who work for different Brazilian companies. %The survey asks about a set of coding practices that may lead testers and programmers to insert smells in the tests they design and code. We analyzed programming practices in the creation of test cases and the presence of test smells during test execution, from the analysis of the frequency of tests with design patterns addressed in this study. 
Our analysis may provide insights towards a better understanding \textcolor{purple}{of }%on
how and which practices may lead to the insertion of test smells in test code. 
 
% ---isso aqui é conclusão
% In this paper we show that independent of experience time, professionals agree they commonly use practices that contribute to the generation of test smells.

%The study carried out an expert survey  to investigate how the programming practices contribute to the insertion of test smells in test code. 
The following research questions were addressed in this study:
% \vspace{-2pt}
\begin{enumerate}[label=\bf RQ\arabic*: ,leftmargin=1.6cm]
    \item \textbf{Do professionals use test case design practices that might lead to test smell insertion?} We investigate whether bad design choices may be related to test smells.
    
    \item \textbf{Does the professional experience interfere with test smell insertion?} We investigate whether, over time, professionals improve the test creation process.
    
    \item \textbf{Which are the practices present in professionals' daily activities that lead test smells insertion?} We investigate which test smells are associated with the most frequent professionals' practices.
\end{enumerate}

%estrutura das demais seções

%The remainder of this paper is structured as follows: Section \ref{back} introduces the concept of test smells; Section \ref{methodology} details the research methodology employed in this study; Section \ref{results} presents the survey's results; Section \ref{discussion} discusses the main findings of this investigation; Section \ref{relatedWork} discusses related work; Section \ref{threats} presents the threats to validity, and Section \ref{conclusion} draws concluding remarks.
\section{Test Smells}

\label{back}

%Software testing is an activity in which systems or components are executed under specific conditions, the results are analyzed, and an evaluation of identified aspect found is made \cite{ieee8292008}. %Usually, software testing is performed from requirement phase to software maintenance and evolution. In general, it aims to guarantee software quality by reducing the risk of failure. 

%Software testing counts on \textit{manually} \textit{automatically} executed tests.%, in which automated tools are used to run scripts (test codes) over system code. 
Automated tests may yield more efficient results when compared to manually executed ones. Due to their capability of being easily repeated several times, and the lack of human interference, automated tests might lead to reductions in both time and execution effort~\cite{Garousi2018mapping,yusifouglu2015mapping}. However, developing test code is not a trivial task, and once it has been poorly designed, the use of automated tools may not ensure the quality of the system~\cite{Palomba2016testsmells,virginio2019coverage}.  As aforementioned, in real-world practice, developers are likely to use anti-patterns during test creation and evolution, which %\textcolor{red}{\sout{what}} 
may lead to mistakes in test code implementation~\cite{Bavota2012smells,Deursen2001refactoring}. These anti-patterns may negatively impact the maintenance of the test code \cite{Rompaey2006ArgoUML}. %Tests code with many test smells may become complex, and hard to understand and modify. Therefore, test smells harm important characteristics of tests, such as repeatability, independence, and stability \cite{Rompaey2006ArgoUML}.

Several studies have investigated different types of smells. Initially, Deursen et al. \cite{Deursen2001refactoring} defined a catalog of 11 test smells and refactorings to remove test smells from test code. After that, many other authors extended this catalog and analyzed the %smells 
effects of the smells on both production and test code 
\cite{Bavota2012smells,bavota2015test,Garousi2016STCE,Greiler2013testsmells,Meszaros2003,Palomba2016testsmells,palomba2019,peruma2018smell,Deursen2001refactoring,Rompaey2006ArgoUML,virginio2019coverage}.
For example, Garousi and Küçük \cite{Garousi2018mapping} found more than 190 test smells on a literature review of 166 studies. %Each study examined from 1 to 86 test smells. 

In this study, we selected 14 types of test smells, which are frequently studied and implemented in cutting-edge test smell detection tools \cite{Meszaros2003,peruma2018smell,Deursen2001refactoring}. These are described next:

\begin{itemize}
\item \textbf{Assertion Roulette (AR)}. It occurs when tests present assertions with no explanation in test methods. If one of those assertions fails, it is not possible to identify which one caused the problem;

\item \textbf{Conditional Test Logic (CTL)}. It occurs when tests present conditional logic (if-else or repeat instructions). Tests with this structure do not guarantee that the same flow is verified, besides allowing a particular piece of code not to be tested; 

\item \textbf{Constructor Initialization (CI)}. This smell occurs when test methods present a constructor;

\item \textbf{Eager Test (ET)}. It occurs when a test method checks many object methods at the same time. This test may be hard to understand, and to execute; 

\item \textbf{Empty Test (EpT)}. Occurs when test methods do not contain executable assertions;

\item \textbf{For Testers Only (FTO)}. Occurs when a production class has methods only used by test methods;

\item \textbf{General Fixture (GF)}. It occurs when the configuration file is generic, and different tests perform tests using part of configuration data. Those files may be hard to read, understand, and they may slower test execution;

\item \textbf{Indirect Testing (IT)}. It occurs when test class has methods that perform tests in different objects because there are references to those objects at the test class, for example.

\item \textbf{Magic Numbers (MN)}. Occurs when tests present a literal number as a test parameter;

\item \textbf{Mystery Guest (MG)}. This smell occurs when the test uses an external resource, such as a file with test data. If the external file is removed, the test results may fail;

\item \textbf{Redundant Print (RP)}. Occurs when test methods contain irrelevant print statements;

\item \textbf{Resource Optimism (RO)}. It occurs when the test code contains optimist assumptions about the presence or absence of external resources. The test may return a positive result once, but it may fail for the next times;

\item \textbf{Test Code Duplication (TCD)}. Occurs when test code has undesired duplication; 

\item \textbf{Test Run War (TRW)}. It occurs when a test passes only when it is performed isolated and fails when it is performed with another test at the same time.

\end{itemize}

%Since testing is an important activity into the process of software development, 

%It is of utmost importance to properly design test code and ensure its quality. Therefore, the analysis of professionals programming practices can contribute to understand how smells are commonly inserted into the test code and to improving the creation and maintenance of test code. 
\section{Research Methodology}
\label{methodology}
%\textcolor{blue}{This session presents the research methodology including design, pilot and data analysis.}
%This study aims to understand whether software engineers non-intentionally insert test smells when developing test code. 
%This study carried out an expert survey \cite{Linaker:TR:2015} to investigate how the programming practices contribute to the insertion of test smells in test code. 
% and also to describe and compare the presence of different smells \cite{pfleeger2001principles1}.
%We defined the following research questions:

%\begin{enumerate}[label=\bf RQ\arabic*: ,leftmargin=1.1cm]
%    \item \textbf{Do professionals use test case design practices that might lead to test smell insertion?} We investigate whether bad design choices may be related to test smells. %Bavota et al. \cite{Bavota2012smells} analyzed 637 manually created JUnit tests and     found test smells on 82\% of them. 
    % The test code was manually created.
    % and may indicate that professionals intentionally adopt bad codding choices.
    
%    \item \textbf{Does the professional experience interfere with test smell insertion?} We investigate whether, over time, professionals improve the test creation process.
    
%    \item \textbf{Which are the practices present in professionals' daily activities that lead test smells insertion?} We investigate which test smells are associated to the most frequently professionals practices.
%\end{enumerate}

\subsection{Survey Design}
For this study, we used the design of observation by case control. It is a descriptive design used to investigate previous situations to support the understanding \textcolor{purple}{of} a current phenomenon \cite{kitchenham2002principles2}. We are interested in detecting the most common practices that may introduce test smells.
%For this study, we use the design of observation by case control to identify the behavior of professionals \cite{kitchenham2002principles2}. We are interested in detecting the most common anti-patterns.
%Our study investigates the most cited test smells in the literature.
% \cite{Bavota2012smells,bavota2015test,Garousi2016STCE,Greiler2013testsmells,Meszaros2003,Palomba2016testsmells,palomba2019,peruma2018smell,Deursen2001refactoring,Rompaey2006ArgoUML,virginio2019coverage}.
In the questionnaire, we did not use the \textit{test smell} term or any of its synonymous not to influence the respondents. Instead, we transcribed the rationale for each test smell and converted them into practices for both test creation and execution. Once the respondents claim they commonly adopt a given practice, and the practice is the definition of a test smell, they assume that they either insert (during test creation) or identify (during test execution) a test smell. Table \ref{table.questions} shows examples of those practices.%\looseness=-1

%Once a respondent claims she commonly adopt a given practice, and the practice is the definition of a test smell, she assumes that she either insert (during test creation) or identify (during test execution) a test smell. Table \ref{table.questions} shows examples of those practices.%\looseness=-1 

\begin{table*}[t]
% \vspace{-14pt}
\setlength{\tabcolsep}{0.5em}
    \def \arraystretch{1.4}
    \centering
    \footnotesize
    \caption{Examples of practices related to test smells.}
    \label{table.questions}
    \begin{tabular}{m{3cm}m{4.3cm}m{4.1cm}}
% Creation and Execution of Test Smells
    \toprule
    \textbf{Test Smell} & \textbf{Practices for Test Creation} & \textbf{Practices for Test Execution} \\
    \midrule
    
    \textbf{Mystery Guest} & I often create test cases using some configuration file (or supplementary) as support & A test case fails due to unavailability of access to some configuration file.\\

% Resource Optimism & ``When creating a test, I analyze whether it can be run at the same time with others or whether it should be performed in isolation, due to the availability of external resources.'' & ``A test case fails because of unavailability of access to some external resource."\\   

%  Test Run War & ``I analyze the possibility of one test failing to use one feature that is being used at the same time by another test.'' & ``Repeat some test case by previously failing due to competition with some other test case that was running at the same time."\\
 
    \textbf{Eager Test} & I often create tests with a high number of parameters (number of files, database record, etc.). & I run some tests without understanding what its purpose is.\\
 
    \textbf{Assertion Roulette} & I pack different test cases into one (i.e., put together tests that could be run separately). & Some tests fail and it is not possible to identify the failure cause.\\

%  Indirect Test & ``I create tests that rely on features that may not have their own validation test (e.g. a test that involves retrieval of database information, but there is no test to validate database search).'' & ``Run a test that depends on some external resource that does not have test for direct validation."\\
 
    \textbf{For Testers Only} & I have already created a test to validate some feature that will not be used in the production environment & I run some tests to validate features that will not be used in the production environment.\\
 
%  Magic Number & ``I have already created a test with a high value for a specific parameter (e.g. number of records in the database, number of files in the folder) even though it makes it difficult to repeat them.'' & ``Run test with a high value for a specific parameter (ex: number of records in the database, number of files in folder) even though it makes it difficult to repeat them."\\
 
    \textbf{Conditional Test Logic} & I have already created conditional or repeating tests. & I run tests with conditional or repeating structure.\\
    
    \textbf{Empty Test} & I have already created an empty test, with no executable statement. & I find empty test, with no executable statement.\\
 
%  General Fixture & ``I usually open a data file from a configuration file.'' & ``Perform tests that could be run faster (maintaining the test goal) by modifying the contents of the configuration file."\\
 
%  Redundant Print & ``I often create test with printing or output rendering redundantly, or unnecessarily.'' & ``Run test with printout or result display redundantly, or unnecessarily."\\
 
%  Constructor Initialization & ``I have already created a test considering the existence of a resource, without checking the existence or availability of it.'' & ``Run test considering the existence of a resource, without checking the existence or availability of the resource."\\
 
% Test Code Duplication & ``Is there verification for detection of duplicate tests (with the same writing or with different writing and same purpose)?'' & ``Find duplicate test (with the same writing or different writing)."\\
\bottomrule
\end{tabular}
\vspace{-10pt}
\end{table*}    
    
The questionnaire was split into three blocks. The first one leverages the respondents' profile. It encompasses thirteen questions, aimed to identify respondents' age, gender, degree, and testing and programming skills. The second one encompasses fourteen statements and six complementary questions: four categorical objective questions and two open-ended questions. The statements describe practices related to test smells. The respondents chose one out of five possible answers (\textit{always, frequently, rarely, never}, and \textit{not applicable}) for each statement, where \textit{always} indicates the adoption of bad practice for test creation. For example, the statement \textit{``I've already created a test to validate a feature that would not be used in the production environment''} corresponds to the \texttt{For testers only} smell. Whether the answer was \textit{``Always''}, it meant the respondent usually uses that practice in his daily tasks, and therefore it is likely he commonly inserts this smell in the test code he develops.
% introduce \texttt{For testers only} smell.
% 02, ie a high response rate ``Always'' would indicate the adoption of good practices in the creation of the test and the others in a negative way, in this case, a high response rate ``Always'' would indicate the adoption of bad practices to find out if respondents have the habit of using them. 
~The six complementary questions are designed to understand how the professionals %are treated during
deal with the test creation process.

The third block is similar to the second one. However, there are fourteen statements taking the perspective of who executes the tests, in which a high response rate (i.e., \textit{Always}) indicates that a respondent usually come across tests with test smells; besides, there is one complementary open-ended question, aimed to understand which problems the professionals are likely to deal with when executing the tests. 
%For this last block, each affirmative has a negative perspective, in which a high response rate (\textit{always}) indicates they face test with test smells.\footnote{The survey answers are available at \url{https://bit.ly/33ZOESi}} 

% \subsection{Pilot Study}
It is worth mentioning that we executed a pilot survey with 4 professionals to identify improvement opportunities. Based on their responses, we could improve the questionnaire prior to running the actual survey.
%\We ran a pilot survey with 4 professionals from 3 different companies to identify improvement opportunities. Based on their responses, we could improve the questionnaire.% so as to provide better understanding of the respondents.

The questionnaire used in this survey is available online\footnote{A copy of the questionnaire is available online at \url{https://bit.ly/2RQVDdc}}. Data were gathered from April 3rd to June 3rd in 2019.

\subsection{Data analysis}
% Before the start of the checks, the frequencies for the responses involving the ``Resource Optimism'' and ``Test Run War'' tests were reversed so that the analyzes were made based on the absence of use of good practices and the full use of good practices.

To answer RQ1, we analyzed the objective questions about test creation and test execution related to test smells. %Those questions had 5 possible answers: ``Always'', ``Frequently'', ``Rarely'', ``Never'', and ``Not applicable''. 
~Each question encompassed an affirmative statement aimed to describe daily practices in software testing, such as \textit{``I usually found empty test''} and \textit{``I usually create test with part of external configuration file''}. 
% The frequency of usage of those practices and presence of the daily situations were presented by the proportion of responses associated to each test smell, because they involve nominal data \cite{kitchenham2003principles6}. 

For RQ2, we compared the professional experience with the frequency of the use of test smells. 
% The years of experience were grouped into 2 years scale and the proportion of usage of
We also used the same answer format as RQ1, but only considered
% We considered the respondents professional activities according their experience and the questions focused on 
the test creation process. During the test execution process, professionals identify test smells instead of creating them. %We aim to investigate whether there is a trend of behavior change with increasing work experience.
% For example, analyzing the frequency of \textit{Always} and \textit{Frequently} responses from less experienced to more experienced professionals, we may identify a rate reduction \textcolor{red}{UMA TAXA DE REDUÇÃO DE QUE??}. This behavior would indicate an upward trend in the use of good practice with increasing professional experience. We may also find an increased use of the ``Never'' answer. In this case, the other frequencies should also be analyzed so that it is possible to identify if there is any tendency of behavior change with the increase of professional experience. %\textcolor{red}{for example, ``colocar um exemplo de questão aqui'' e ``outro aqui"}.
% answers ``Always'', ``Frequently', ``Rarely'', ``Never'', and ``Not applicable'' for test creation and test execution process had been analyzed. 
% We also considered the professional activities of respondents across of years.

To answer RQ3, we grouped the practices by frequency to identify which ones are most commonly used. The practices may be associated to one or more test smells according to their characteristics, such as external file usage, conditional structure, and programming style.

The three open questions were analyzed through coding and continuous comparison \cite{kitchenham2015evidence}. Our main intention was to understand the reasons that might lead the software engineers to use practices that may insert test smells. Besides, we were also intended to understand which difficulties they come across when creating and executing tests. The coding task was performed by two researchers and validated by consensus. We also associated some of the practices with the test code characteristics defined by Meszaros et al.~\cite{Meszaros2003}.

Since the questionnaire encompasses both close-ended and open-ended questions, after data collection, we employed open coding to identify additional reasons on why professionals use bad practices on testing activities. Codes were peer-reviewed and changed upon agreement with the authors of this paper. Considering that open questions were optional, we used coding to complement our results. \looseness=-1

\section{Results}
\label{results}

\subsection{Participants}

We sent the invitation to professionals from eight Brazilian companies in a convenience sampling. We created one copy of the questionnaire for each of these companies (C1-C8) and sent them by email.
The different versions of the questionnaire served to control the number of respondents from each company.
Those companies have from 4 to 66 testing professionals, who perform both manual and automated tests, as Table \ref{tab:respondents} shows. 

Besides, we also sent the questionnaire by direct message (D1) and posted it on a Facebook group dedicated to discussing software testing (G1). In total, 305 professionals were contacted.%, by non-probabilistic for convenience and similar snowball sampling technique \cite{kitchenham2015evidence}. 
%~From this set, we received 60 answers (19,67\%).%, but only 26 respondents answered the survey (8,52\%). However, the descriptive questions have a complementary character, so all 60 answers were considered.

As a result, we received 60 answers\footnote{Raw data with all the answers (PT-BR) are available at \url{https://bit.ly/33ZOESi}} (out of 305 potential respondents) from three different states in Brazil: BA (66.7\%), SP (31.7\%), and PR (1.6\%). The respondents were from 22 to 41 years old, and their experience with quality assurance ranged from 0 to 13 years (5.16 on average). The experience as software developers also ranged from 0 to 13 years (average 1.67). Most of the respondents were male (65\%), but results also showed female (32\%) and non-binary gender (3\%). Most of them hold a degree in Computer Science-related courses (83.3\%), some of them hold a degree in other STEM courses (10\%), and some of them hold a degree in other areas (6.7\%). 90\% of the respondents pursued higher education degrees, as follows: 66.7\% hold a bachelor's degree, 21.7\% hold a graduate degree, and 1.6\% hold a postdoc.

%\begin{table}[t]
%\centering
%\begin{small}
%\caption{respondents} \label{tab:respondents}
%\begin{tabular}{p{1.0cm}p{2.0cm}p{1.0cm}}
%% \hline
%Source      &       Employees /\mbox{Members}   &   Answers\\
%\hline
%C1  &   66  &   14\\
%C2  &   30  &    1\\
%C3  &   10  &    0\\
%C4  &    6  &    0\\
%C5  &    5  &    0\\
%C6  &    4  &    4\\
%C7  &    4  &    4\\
%C8  &    4  &    0\\
%D1  &   52  &   35\\
%G1  &  124  &    2\\
%\hline
%\end{tabular}
%\end{small}
%\end{table}

%Acho que a tabela pode ser assim, para economizar espaço
\begin{table}[t]
\centering
\footnotesize
\caption{respondents} 
\label{tab:respondents}
\begin{tabular}{ccc|ccc}
\toprule
\bf Source & \bf Professionals & \bf Answers & \bf Source & \bf Professionals & \bf Answers \\
\midrule
C1  &   66  &   14 & C6  &    4  &    4\\
C2  &   30  &    1 & C7  &    4  &    4\\
C3  &   10  &    0 & C8  &    4  &    0\\
C4  &    6  &    0 & D1  &   52  &   35\\
C5  &    5  &    0 & G1  &  124  &    2\\
\bottomrule
\end{tabular}
\end{table}

Regarding their commonly performed software testing tasks, most of the respondents reported creating and running tests at the same rate (43.3\%); but also executed tests with more frequency than they created (21.7\%); and created tests with more frequency than they executed (13.3\%). Moreover, some respondents only executed test cases (20\%); and other respondents only created them (1.7\%). They performed tests over many different platforms, and most of them (53\%) worked with two or more platforms: Web (65\%), Android (58\%), Desktop (48\%), and Apple (28\%) applications. Backend, microservices, API, mainframe, and cable TV had 1.67\% each. \looseness=-1

Usually, professionals tested solutions for mobile device applications (65\%) and web applications (60\%). We also identified the following domains they work with:
% but also other domains, 
% such as 
embedded systems (23,33\%), cloud computing (18,33\%), information security (11,67\%), internet of things (6,67\%), big data, retail, artificial intelligence, cable tv, bioinformatics, commercial information, desktop system, and payment solutions (1,67\% each one).
% For most of respondents, 
The sources of test creation were
% are 
requirements (48\%); developed software (45\%); source code and use cases (2\% each); and 3\% did not know the source.
% . There are also respondents who are unaware of test code source (3\%). 

% \vspace{0.3cm}\noindent
\subsection{Test creation and execution practices} 

\begin{figure}[t]
    \begin{center}
        \includegraphics[scale=0.59]{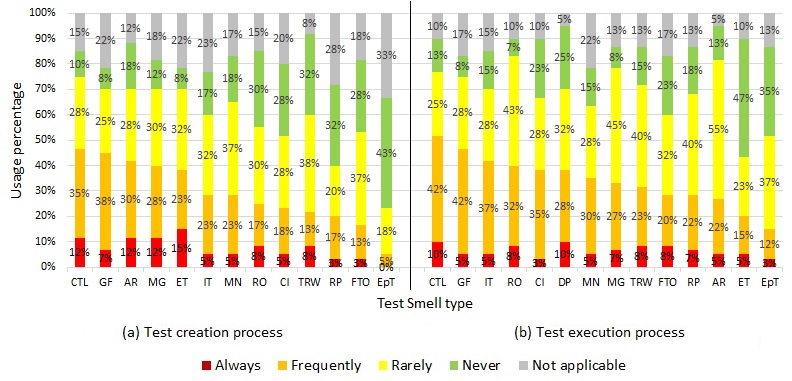}
        \caption{Test Smells frequencies on test creation and execution.}
        \label{fig:testfrequencies}
    \end{center}
    \vspace{-20pt}
\end{figure}

Gathered data made it possible to establish a relationship between test creation and execution practices and the occurrence of test smells. Figure \ref{fig:testfrequencies} summarizes such a relationship, by showing the frequency of test smells usage during (a) test creation and (b) test execution processes, based on the responses. We next discuss the main results. 

The search for test duplication was a personal practice for most of the respondents (48\%). In some cases, it was also a company practice (18\%), or even both, respondent and company practices (6\%). However, for 28\% of respondents, this activity was not applied. Checking tests with the same objective reduces the 
% test smell type, the 
\textbf{Test Duplication} smell.

During the test creation process, \textbf{Conditional Test Logic} and \textbf{General Fixture} were the most frequent test smells. 
% practices directly associated to test smells,
% % described as practices, 
% such as \textit{test creation with generic configuration data} \textcolor{purple}{related to \texttt{General Fixture (GF)} smell} and \textit{creation of test with conditional or repetition structure} \textcolor{purple}{associated to \texttt{Conditional Test Logic (CTL)} smell} were 
% % informed as 
% frequent to most of respondents. 
The former obtained 47\% of \textit{Always} and \textit{Frequently} answers, and the latter, 45\% in both answers, as Figure \ref{fig:testfrequencies}(a) shows. A high rate of those responses may indicate a common use of practices related to those smells.
% Those practices are directly associated to test smells, as Table \ref{table.questions} shows.
%Figure~\ref{fig:testfrequencies} shows the frequency of test smells usage during test creation and execution.
%\textcolor{orange}{For example, during test creation (Figure \ref{fig:testfrequencies}(a)), the bad practices related to Conditional Test Logic were the most reported (12\% of absence and 35\% of rare use of good practices).}
% , in which we highlight:
% Conditional Test Logic (47\%) and General Fixture (45\%).
We also analyzed why developers create tests with bad practices (one open non-mandatory question answered by 45\% of respondents).
% They also could explain why they created tests with this structure, as an open non-mandatory question. It was answered for 45\% of respondents. 
The main reasons are related to
% Among the answers,
company or personally employed standards, limited time, and attempt to reach better coverage and efficiency.
% are the main reasons for using such practices.
%  Among the answers, apply company standard (25.9\%), personal knowledge (18.5\%), best use of time (11.1\%), and to try to reach best coverage and efficiency (11.1\%) are the main reasons for using such practices.

% \begin{figure}[t]
%     \begin{center}
%         \includegraphics[scale=0.9]{images/grap_creation.png}
%         \caption{Test Smells frequency on test creation.}
%         \label{fig:testcreation}
%     \end{center}
% \end{figure}

We also asked whether they have already changed existing test sets because they found the bad practices we presented. 12\% of respondents always perform any changes in the test code, when they find test smells-related practices; 38\% frequently change them; 27\% rarely; 12\% never edit test code, and 11\% answered as not applicable. Among the reasons to change the test, they informed that changing it reduces ambiguities (30\%); improves execution speed (27\%); they need to adequate it to company standards (23\%); they change it when they do not understand the test purpose (13\%); and when the corresponding production class evolves (7\%). 
% and detected no more defects (7\%) due to test code evolution (7\%). \textcolor{red}{eu coloquei de volta pq era dessa forma que estava a alternativa no survey. Se não tiver problema mudar a forma que está escrito, remove. Se precisar manter é só tirar o percentual do final do parágrafo.}
% \textcolor{red}{and detected no more defects (7\%)}.

Furthermore, the respondents had to point out which test structure problems they face. The results indicated that some tests depended on third party resources (52\%); were hard to understand (48\%); contained either unnecessary information or were ambiguous (40\% each); depended on external files (33\%); used external configuration file (10\%); and presented resources limitation (2\%).

Regarding difficulties in creating test cases (one open non-mandatory question answered by 38\% of respondents), requirement issues were the most frequent ones (56\%). Other problems were related to difficulties in performing test code reuse, lack of knowledge, issues in production code, code coverage, test environment problems, and time and resources limitation.
% as another complications faced to create test cases.
%To conclude the sequence of questions about test creation, the respondents could inform what difficulties they faced to create test cases, also as an open non-mandatory question, had been answered for 38\%. In the majority of answers, requirement issues are indicated as main reason. Non-clear requirements (39\%), absence of requirement document (30\%) and gap to update this document (9\%) were the related subjects for requirement issues. They also informed to generalize test for reuse of test code (13\%), gap of knowledge (9\%), production code issues, code coverage, test environment problems,  time and resources limitation (4\% each one) as another complications faced to create test cases.

% \subsection{Practices related to tests execution} 
% Similar to the block of questions about test creation and present in the Table \ref{table.questions}, 
The questions regarding test execution also presented a sequence of statements about ordinary situations the developers usually face, in which respondents should answer according to frequency. \textbf{Conditional Test Logic} (52\%) and \textbf{General Fixture} (47\%) were the test smells most cited during test execution, as Figure \ref{fig:testfrequencies}(b) shows.

% \begin{figure}[htbp]
%     \begin{center}
%         \includegraphics[scale=0.9]{images/grap_execution.png}
%         \caption{Test Smells frequency on test execution.}
%         \label{fig:testexecution}
%     \end{center}
% \end{figure}

The most cited problem related to test execution refers to difficulties involving the test environment (34\%). Some of the cited issues were related to the unavailability of the test environment, the demand for 3rd party features, or even low-performance environments. The second most common problem is associated with understanding the test purpose (28\%). They informed that tests are poorly written, which allow multiple interpretations, besides lack of writing standard.
Lack of test maintenance was the third problem (24\%), which involves outdated and incomplete tests due to the evolution of the system code.

%The most cited problem related to test execution refers to difficulties to understand the test purpose (31\%). They informed that tests are poorly written, which allow multiple interpretations, besides lack writing standard. The second most common problem is associated with the availability of the test environment (28\%). Some of the cited issues were related to unavailability of test environment and 3rd party feature, and low performance environments. Lack of test maintenance was the third problem (21\%), which involves outdated and incomplete tests due to evolution of system code.

\subsection{Professional Experience}

Although most respondents reported they create and execute
tests at the same proportion, our investigation presents a different scenario as the tester gets more experienced. 
For the first two years of experience, there was a predominance of test execution. 64\% only executed tests; 9\% frequently executed tests and occasionally created them, and 27\% created and executed them at the same proportion.
Over the years, mainly after 10 years of experience, practitioners tend to create more tests than execute them. We found that the ones who only created tests had at least 12 years of professional experience.
Thus, the less experienced the tester, the less number of tests they create and the more tests they execute.
%she creates and the more number of tests she executes.
% \begin{figure}[htbp]
%     \begin{center}
%         \includegraphics[scale=0.85]{images/experience_01.png}
%         \caption{Daily activities according professional experience.}
%         \label{fig:activitytime}
%     \end{center}
% \end{figure}

\begin{figure}[t]
    \begin{center}
        \includegraphics[scale=0.41]{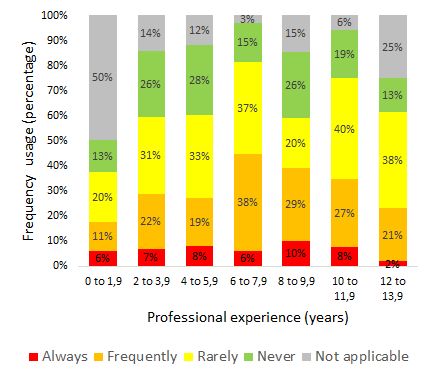}
        \caption{Test Smells frequencies on test creation according professional experience.}
        \label{fig:xpfrequencies}
    \end{center}
    \vspace{-20pt}
\end{figure}

We also analyzed the use of good practices to create tests as professionals become more experienced.
During the creation of a test, the use of good practices tends to increase over the years, as \textit{Rarely} and \textit{Never} got more responses than the remainder. However, gathered data showed a slight reduction between eight and ten years, but it grows back from ten years of experience. On the other hand, the use of bad practices also increases over the years, with \textit{Always} and \textit{Frequently} gaining more responses, to respondents with up to eight years of experience, and decreasing thereafter, as Figure \ref{fig:xpfrequencies} shows. %\looseness=-1

\section{Discussion}
\label{discussion}

%Along this section, we discuss the main findings regarding each RQ.

\subsection{Test design practices that lead to test smells insertion (RQ1)}

We observed that some practices might lead to the introduction of test smells in test code. Therefore, we analyzed the frequency of those practices for test creation and execution. 

In the former, we observed that every test smell presented at least three out of four possible answers (\textit{Always, Frequently, Rarely, and Never}). We then classified such gathered data into two groups: Commonly-used practices group (CPG) and the Unused practices group (UPG). CPG contains test smells that mostly present \textit{Always} and \textit{Frequently} as answers, and UPG that mostly present \textit{Rarely} and \textit{Never} as answers. We consider a test smell belonging to one of the groups when the difference between the rates of \textit{Always} and \textit{Frequently} versus \textit{Rarely} and \textit{Never} is greater than 10\%. For example, \texttt{Empty Test, For testers only, Test run War, Constructor Initialization, Resource Optimism, Redundant Print, Magic Number, Indirect test} belong to the UPG group, which means professionals rarely insert those smells on testing activities. Conversely, professionals frequently adopt practices related to \texttt{General Fixture}, the only member of the CPG group. Thus, it is possible that they create many tests with that smell, which may compromise test maintenance \cite{tufano2016}. Still, four test smells presented a similar frequency of pertinence to both groups (less than 10\% of difference). For them, there was not a pattern among respondents. For instance, the \texttt{Eager Test} smell obtained 38\% to CPG and 40\% to UPG.

In the latter, UPG contains the following test smells: 
\texttt{Empty Test, Eager Test, Assertion Roulette, Redundant Print, Duplicated Test, Test Run War, For testers Only, Mystery Guest, Constructor Initialization, and Resource Optimism}, which means that professionals rarely face those smells during testing. On the other hand, professionals frequently find practices related to two test smells, which are part of CPG: \texttt{General Fixture} and \texttt{Conditional Test Logic}. Besides, other two test smells presented a similar frequency of pertinence to both groups, \texttt{Indirect Test} and \texttt{Magic Number}. Thus, there is no perceived standard among respondents for them. In general, our study identified that all test smells we analyzed appear in the testing activities. They all were cited by respondents, even rarely.

We also analyzed the reasons that lead professionals to adopt the practices presented in the survey. Since this data came from open-ended questions, we identified 16 different codes, in which the most common ones are: \textit{company standard, personal standard, project politics, professional experience, saving time} and \textit{improving coverage}. For example, professional \#26 reported applying company standards when creating tests, and those standards may insert smells. He said he commonly applies bad practices \textit{``to match company development standards.''} Otherwise, professional \#54 reported personal standard when said that \textit {``I group tests by module so that tests can be executed sequentially without compromising effectiveness.''} This behavior also indicates that participants may have misunderstood the definition of test smells. When grouping tests, it is possible to insert the \texttt{Assertion Roulette} test smell and to compromise test independence. Similar situation occurred to professionals \#14, \#16, \#27, \#50 and \#59. 

Table \ref{table.answers} summarizes the answers for each RQ.
%The following frame synthesizes the findings supporting answering RQ1:

%\begin{framed}
%    \noindent
%      We found that professionals do adopt practices for test case design, which introduce test smells. Usually, those practices come from improper personal and company standards.
%\end{framed}

\begin{table*}[t]
    \def \arraystretch{1.4}
    \centering
    \footnotesize
    \caption{Summary of answers for each research question.}
    \label{table.answers}
    \begin{tabular}{m{1.0cm}m{11.0cm}}
    \toprule
    \textbf{RQ} & \textbf{Answer} \\
    \midrule
    
    \textbf{RQ1} & We found that professionals do adopt practices for test case design, which introduce test smells. Usually, those practices come from improper personal and company standards.\\

    \textbf{RQ2} & Experienced software testing professionals may not produce less test smells than inexperienced ones.\\
 
    \textbf{RQ3} & The practices most present in the daily life of professionals that lead to test smells insertion were the  use of conditional structure or repetition and the use of generic configuration data.\\
\bottomrule
\end{tabular}
\end{table*}

\subsection{Professional experience and its interference on test smell insertion (RQ2)}

Although we analyzed the experience of practitioners and its influence on the adoption of practices that introduce test smells, we have not identified any clear correlation. 
For example, the \textit{Always} option indicates they always use bad practices. Regarding the test creation process, we cannot infer that inexperienced professionals introduce more smells on tests than the experienced ones. We found the following rates (from less experienced to the most experienced professionals): 6\% of professionals from 0-2 work years answered \textit{Always}, 7\% to 2-4 work years, 8\% to 4-6, 6\% to 6-8, 10\% to 8-10, 8\% 10-12 and 2\% for over 10 years. 

When testers are inexperienced programmers, they may write lower quality tests, but when they are more experienced, they can carry programming biases that may also contain bad practices.
The absence of a tendency indicates a non-behavioral change between less and more experienced software testing professionals. 
%It answers to RQ2, avalilable in Table \ref{table.answers}.}
%The following frame synthesizes the findings supporting answering RQ2:

%\begin{framed}
%    \noindent
%        Experienced \textcolor{blue}{testing} professionals may not produce less test smells than inexperienced ones.
%\end{framed}

\subsection{Professional practices that might lead to introducing test smells (RQ3)}

Tests can be performed either manually or automatically. Although there are specific tools to support test automation \cite{Fraser2011Evosuite,smeets2009automated}, 62\% of respondents perform more manual than automated tests. In addition, they are also
inexperienced in software development; 55\% have no experience with software development (less than 2 years of experience on average). Lack of experience with software development in general may contribute to the use of bad practices. %\textcolor{blue}{We believe that experience with software code development may contribute to the use of good programming practices, and it may influence on test code quality.}
% We believe these factors favor the smells insertion, as they allow professionals to intentionally use practices that introduce smells into the test code.

According to the practices explored in this study, we identified that two development activities are very present in professionals' daily life:
\textit{(ii) the use of generic configuration data}, which produces \texttt{General Fixture}.  This smell is the most frequent on test creation and execution processes for CPG;
and \textit{(i) the use of conditional structure or repetition}, which is directly associated to \texttt{Conditional Test logic} smell. It was the second most detected smell on test execution (CPG group).

Professionals indicated they commonly face several problems with tests, such as poorly written tests and outdated and incomplete test procedures. According to them, when tests are associated with generic configuration data, test case gets hard to understand and also may cause incorrect results due to the lack of maintenance. Moreover, the presence of conditional logic on tests does not make it clear which structure of the production code is being covered. Understanding which practices are most prevalent in the professionals' activities supports improving test quality. Other problems are related to incomplete, outdated, or lack of documentation, which make difficult to reach traceability, evolution, and maintenance testing tasks.
%\textcolor{blue}{It supports to answer to RQ3, available in Table \ref{table.answers}.}
%The following frame synthesizes the findings supporting answering RQ3:

%\begin{framed}
%    \noindent
%    The practices most present \textcolor{blue}{in the daily life of professionals} that lead to test smells insertion were the  use of conditional structure or repetition and the use of generic configuration data.
%\end{framed}

\section{Threats to validity}
\label{threats}

\textbf{Internal validity}. Although there are more than 100 test smells, this study only considered 14 of them. However, we selected the most frequently smells discussed in the literature. In addition, the smells were presented in the survey as practices. To mitigate ambiguities and text comprehension, we applied a pilot with 4 testers from different companies. 

\noindent
\textbf{External validity}. We sent the survey to 305 professionals, but only received 60 answers. Although our results may not generalize, they provide an initial view of the practices adopted by testers. 
Despite the limitations, we performed the survey procedure and data was made available to allow further replications of this study.

\noindent
\textbf{Construct validity}. During the survey, it was not informed that the questions referred to test smells to do not influence the results. For open questions, a peer-reviewed coding process was performed to avoid bias. The survey and answers were written in Portuguese and translated into English by one author, but reviewed by all others. 

\noindent
\textbf{Conclusion validity.} Since this is a qualitative study, we cannot use a statistical argument to generalize the results. We mitigated this threat by sending the questionnaire to different companies and states of Brazil. %However, we got answers from only three of them, with more concentration in Bahia.

\section{Related work}
\label{relatedWork}

Bavota et al. \cite{bavota2015test} presented a case study to investigate the impact of test smells on maintenance activities. In that study, developers and students analyzed testing code to compare whether their experience would make a difference in identifying test smells. As a result, they identified that test smells have a significantly negative impact on maintenance activities. 
Conversely, our study identified for the professionals that we surveyed, the experience does not interfere in the test smell introduction during the creation and execution of a test.

Palomba and Zaidman \cite{palomba2019} conducted a study to analyze the relationship between test smells and flaky tests. They analyzed test smells identified in a multivocal review and detected five flakiness-inducing test smell types \cite{Garousi2018mapping}. After that, a semi-structured interview was conducted with 10 developers with more than 10 years of experience.
As a result, no new test smell was found as a flakiness-inducing, and the five previously identified test smells were ratified. In our study, we were careful not to use the expression test smell as a means to prevent any influence on the respondents' answers. Besides, professionals with different backgrounds participated in the survey.

Tufano et al. \cite{tufano2016} proposed a study with 19 participants to investigate developers' perception of test smells. They performed an empirical investigation to analyze where test smells occur at source code. The results showed that developers generally do not recognize test smells, and they are usually introduced since the first commit at the repository. Our study found that most of the professionals frequently adopt practices that lead to smells. However, we use the term \textit{practices} instead of \textit{test smells}.
We did not find any study investigating how professional practices affect the introduction of test smells.
%\vspace{-12pt}
\section{Concluding Remarks}
\label{conclusion}

Test smells may decrease the quality and maintenance of the testing code. 
Our study aimed at identifying whether professionals know that they introduce smells during test activities, besides understanding whether professional experience influences the adoption of test smells. Our initial results showed that they commit bad practices out of habit or because they follow company standards. %\texttt{General Fixture} was the most common test smell.
We also found that no bad practice was %completely 
utterly unknown, and all 14 are adopted, even if rarely.

Furthermore, we found that experienced professionals do not insert a few smells than inexperienced ones.
As future work, we intend to extend this study to understand better how the industry deals with test smells.
We also intend to investigate which techniques companies and professionals could adopt to reduce vices that may lead to test smells.

\subsubsection*{Acknowledgments.}
This research was partially funded by INES 2.0; CNPq grants 465614/2014-0 and 408356/2018-9 and FAPESB grant JCB0060/2016.

% \subsubsection*{Acknowledgments.} 
% We would like to thank the 60 anonymous respondents who participated in the survey. This study was financed in part by the Coordenação de Aperfeiçoamento de Pessoal de Nível Superior - Brasil (CAPES) - Finance Code 001. 

\bibliographystyle{splncs04}
\bibliography{ref}

\begin{thebibliography}{10}
\providecommand{\url}[1]{\texttt{#1}}
\providecommand{\urlprefix}{URL }
\providecommand{\doi}[1]{https://doi.org/#1}

\bibitem{Bavota2012smells}
{Bavota}, G., {Qusef}, A., {Oliveto}, R., {Lucia}, A., {Binkley}, D.: An
  empirical analysis of the distribution of unit test smells and their impact
  on software maintenance. In: 28th IEEE International Conference on Software
  Maintenance (ICSM) (2012)

\bibitem{bavota2015test}
Bavota, G., Qusef, A., Oliveto, R., Lucia, A., Binkley, D.: Are test smells
  really harmful? {A}n empirical study. Empirical Software Engineering
  \textbf{20}(4) (2015)

\bibitem{Fraser2011Evosuite}
Fraser, G., Arcuri, A.: Evosuite: Automatic test suite generation for
  object-oriented software. In: 13th European Conference on Foundations of
  Software Engineering. ESEC/FSE, ACM, New York, NY, USA (2011)

\bibitem{Garousi2016STCE}
{Garousi}, V., {Felderer}, M.: Developing, verifying, and maintaining
  high-quality automated test scripts. IEEE Software  \textbf{33}(3) (2016)

\bibitem{Garousi2018mapping}
Garousi, V., K{\"u}{\c{c}}{\"u}k, B.: Smells in software test code: A survey of
  knowledge in industry and academia. Journal of systems and software
  \textbf{138} (2018)

\bibitem{Greiler2013testsmells}
{Greiler}, M., {van Deursen}, A., {Storey}, M.: Automated detection of test
  fixture strategies and smells. In: 2013 IEEE Sixth International Conference
  on Software Testing, Verification and Validation (2013)

\bibitem{kitchenham2002principles2}
Kitchenham, B.A., Pfleeger, S.L.: Principles of survey research part 2:
  designing a survey. ACM SIGSOFT Software Engineering Notes  \textbf{27}(1)
  (2002)

\bibitem{kitchenham2015evidence}
Kitchenham, B.A., Budgen, D., Brereton, P.: Evidence-based software engineering
  and systematic reviews, vol.~4. CRC press (2015)

\bibitem{Linaker:TR:2015}
Linåker, J., Sulaman, S.M., Maiani~de Mello, R., Höst, M.: Guidelines for
  conducting surveys in software engineering. Tech. rep., [Publisher
  information missing] (2015),
  \url{https://lup.lub.lu.se/search/ws/files/6062997/5463412.pdf}

\bibitem{Meszaros2003}
Meszaros, G., Smith, S.M., Andrea, J.: The test automation manifesto. In:
  Maurer, F., Wells, D. (eds.) Extreme Programming and Agile Methods - XP/Agile
  Universe 2003. Springer Berlin Heidelberg (2003)

\bibitem{Palomba2016testsmells}
Palomba, F., Di~Nucci, D., Panichella, A., Oliveto, R., De~Lucia, A.: On the
  diffusion of test smells in automatically generated test code: An empirical
  study. In: {9th International Workshop on Search-based Software Testing}. ACM
  (2016)

\bibitem{palomba2019}
Palomba, F., Zaidman, A.: The smell of fear: On the relation between test
  smells and flaky tests. Empirical Software Engineering  (2019)

\bibitem{peruma2018smell}
Peruma, A.S.A.: What the Smell? An Empirical Investigation on the Distribution
  and Severity of Test Smells in Open Source Android Applications. {PhD}
  {T}hesis, Rochester Institute of Technology (2018)

\bibitem{smeets2009automated}
Smeets, N., Simons, A.J.: Automated unit testing with {R}andoop, {JW}alk and
  $\mu${J}ava versus manual {JU}nit testing. Research report, Department of
  Computer Science, University of Sheffield/University of Antwerp, Sheffield,
  Antwerp (2011)

\bibitem{tufano2016}
Tufano, M., Palomba, F., Bavota, G., Di~Penta, M., Oliveto, R., De~Lucia, A.,
  Poshyvanyk, D.: An empirical investigation into the nature of test smells.
  In: 31st International Conference on Automated Software Engineering. IEEE
  (2016)

\bibitem{Deursen2001refactoring}
Van~Deursen, A., Moonen, L., Van Den~Bergh, A., Kok, G.: Refactoring test code.
  In: Proceedings of the 2nd international conference on extreme programming
  and flexible processes in software engineering (XP) (2001)

\bibitem{Rompaey2006ArgoUML}
Van~Rompaey, B., Du~Bois, B., Demeyer, S.: Characterizing the relative
  significance of a test smell. In: 22nd International Conference on Software
  Maintenance (2006)

\bibitem{virginio2019coverage}
Virg{\'\i}nio, T., Santana, R., Martins, L.A., Soares, L.R., Costa, H.,
  Machado, I.: On the influence of test smells on test coverage. In:
  Proceedings of the XXXIII Brazilian Symposium on Software Engineering. ACM
  (2019)

\bibitem{wiederseiner2010open}
Wiederseiner, C., Jolly, S.A., Garousi, V., Eskandar, M.M.: An open-source tool
  for automated generation of black-box xunit test code and its industrial
  evaluation. In: Bottaci, L., Fraser, G. (eds.) Testing -- Practice and
  Research Techniques. Springer Berlin Heidelberg (2010)

\bibitem{yusifouglu2015mapping}
Yusifoğlu, V.G., Amannejad, Y., Can, A.B.: Software test-code engineering: A
  systematic mapping. Information and Software Technology  \textbf{58} (2015)

\end{thebibliography}

\end{document}